# Interplanetary Coronal Mass Ejections observed by MESSENGER and *Venus Express*

## S. W. Good and R. J. Forsyth

Blackett Laboratory, Imperial College London, London, SW7 2AZ, UK

simon.good07@imperial.ac.uk

**Abstract**

Interplanetary coronal mass ejections (ICMEs) observed by the MESSENGER and *Venus Express* spacecraft have been catalogued and analysed. The ICMEs were identified by a relatively smooth rotation of the magnetic field direction consistent with a flux rope structure, coinciding with a relatively enhanced magnetic field strength. A total of 35 ICMEs were found in the surveyed MESSENGER data (primarily from March 2007 to April 2012), and 84 ICMEs in the surveyed *Venus Express* data (from May 2006 to December 2013). The ICME flux rope configurations have been determined. Ropes with northward leading edges were about four times more common than ropes with southward leading edges, in agreement with a previously established solar cycle dependence. Ropes with low inclinations to the solar equatorial plane were about four times more common than ropes with high inclinations, possibly an observational effect. Left and right-handed ropes were observed in almost equal numbers. In addition, data from MESSENGER, *Venus Express*, STEREO-A, STEREO-B and ACE were examined for multipoint signatures of the catalogued ICMEs. For spacecraft separations below 15° in heliocentric longitude, the second spacecraft observed the ICME flux rope in 82% of cases; this percentage dropped to 49% for separations between 15 and 30°, to 18% for separations between 30 and 45°, and to 12% for separations between 45 and 60°. As spacecraft separation increased, it became increasingly likely that only the sheath and not the flux rope of the ICME was observed, in agreement with the notion that ICME flux ropes are smaller in longitudinal extent than the shocks or discontinuities that they often drive. Furthermore, this study has identified 23 ICMEs observed by pairs of spacecraft close to radial alignment. A detailed analysis of these events could lead to a better understanding of how ICMEs evolve during propagation.

## 1 Introduction

Coronal mass ejections (CMEs) are large scale eruptions of magnetic flux, helicity and plasma from the Sun distinct from the continuous outflow of the solar wind. Their interplanetary manifestations, ICMEs, display a range of signatures when observed by spacecraft *in situ* (Zurbuchen and Richardson, 2006). ICMEs that have low plasma temperatures, enhanced B-field magnitudes, and a B-field direction that rotates smoothly for ~1 day when observed at 1 AU are known as 'magnetic clouds' (Burlaga et al., 1981; Burlaga, 1991). The field rotations within magnetic clouds are consistent with flux rope structures (Goldstein, 1983), which comprise nested, helical field lines of decreasing pitch angle as the central axis of the rope is approached. ICMEs are major drivers of space weather at the Earth.

In this work, we present a catalogue of ICMEs observed by the *MErcury Surface, Space ENvironment, GEochemistry, and Ranging* (MESSENGER; MES) spacecraft, and by *Venus Express* (VEX). Despite their primary focus as planetary missions, both spacecraft spent significant amounts of time in the solar wind, allowing observation of transient solar wind structures such as ICMEs. MES data spanning ~5 years have been examined for ICME signatures, including data from the spacecraft's cruise phase in interplanetary space and the first year it spent orbiting Mercury; similarly, VEX data spanning ~7.5 years have been examined, all of which was gathered while the spacecraft orbited Venus.

Neither spacecraft carried dedicated instruments for analysing the dynamics and composition of the solar wind plasma, and so ICMEs have been identified by their magnetic field signatures alone. The identification criteria have been chosen such that the ICMEs identified may well be magnetic clouds also; the absence of temperature data precludes any definite magnetic cloud identifications.

MESSENGER and *Venus Express* offer the most recent *in situ* observations of ICMEs at sub-1 AU heliocentric distances. Given the lack of plasma and compositional data, a comprehensive analysis of the nature of the ICMEs encountered by MES and VEX is not possible, unlike in the case of the pioneering sub-1 AU *Helios* mission (*e.g.*, Bothmer and Schwenn, 1998; Leitner et al., 2007). However, much may be gleaned from the magnetic field measurements alone. In this work, we classify the flux rope configurations for the catalogued ICMEs, and consider how the occurrence of different configuration types is related to the solar cycle. The frequency at which ICMEs were observed by MES and VEX is also considered in terms of the solar cycle.

Perhaps the greatest value to be obtained from ICME observations by MES and VEX can be found in linking observations of individual events at these spacecraft to observations of the same events at 1 AU spacecraft. Taken together, MESSENGER, *Venus Express*, STEREO-A, STEREO-B and ACE/*Wind* offered unprecedented *in situ* coverage of interplanetary space near the solar equatorial (SE) plane, and unprecedented opportunities for multipoint ICME observations. For each ICME observed by MES and VEX, we have identified signatures of the same event at other spacecraft. This analysis has been used to estimate the probability that two spacecraft will observe the same ICME as a function of the spacecraft's heliocentric longitudinal separation. This analysis will be useful for those seeking to perform studies of ICMEs observed by radially aligned spacecraft involving MES or VEX, or studies that consider longitudinal variations in ICMEs.

Other ICME catalogues have recently been compiled that use MESSENGER and *Venus Express* data. Winslow et al. (2015) list relatively fast ICMEs observed by MES after March 2011 that drove shocks, displayed an enhanced field magnitude and distorted Mercury's magnetosphere. Similarly, Vech et al. (2015) identify ICMEs observed by VEX that produced significant enhancements in the field magnitude of Venus's induced magnetosphere.

In Section 2, descriptions of the two spacecraft, the data used and data availability are presented. The ICME identification criteria are described in Section 3. In Section 4, the ICME catalogue is presented and analysed.

## 2 Spacecraft and data

NASA's MESSENGER spacecraft was launched in August 2004 and, following a cruise phase through interplanetary space of ~6.5 years, entered orbit around Mercury in March 2011. The first mission to visit Mercury since *Mariner 10* in the 1970s, it carried a range of instrumentation for the study of Mercury's chemistry, geology and magnetosphere (Solomon et al., 2001). MESSENGER orbited the planet with a 12-hour period during its primary mission, reducing to an 8-hour period from April 2012 at the start of the first extended mission. Following total depletion of its propellant, MESSENGER impacted the surface of Mercury in April 2015.

This study makes use of magnetic field measurements made by MESSENGER's magnetometer (MAG; Anderson et al., 2007). The data was obtained from the Planetary Plasma Interactions (PPI) node of the Planetary Data System (PDS) archive (ppi.pds.nasa.gov/index.jsp). Data at a time-averaged resolution of 10 seconds have been used; figures in this work displaying MESSENGER MAG measurements show data at this resolution.

The black-dotted line in Figure 1 show the heliocentric distance of MESSENGER with time, where the *x*-axis spans the full duration of the mission. The grey-shaded region in the lower half of the figure before March 2007 (marked with a vertical dotted line) highlights a period of time where data is largely unavailable or suspect. Grey-shaded regions in the lower half of the figure after March 2007 show data-gap intervals during the remainder of the cruise phase.

Figure 1 also shows the interval of time covered by the MES ICME catalogue presented in this work. The catalogue primarily spans the time from when data became more regularly available, in March 2007, to the time, following planetary insertion, when MESSENGER's orbital period at Mercury changed from 12 to 8 hours in April 2012. The signatures used in this study to identify ICMEs (described in Section 3) were often heavily obscured by bow shock and magnetosphere traversals after April 2012 - three magnetosphere traversals per 24 hours from then onwards, compared to two crossings previously - and so data after this time have not been surveyed. Three additional events have been identified outside the primary catalogue window, two in May 2005 (during a brief period when data was available) and one clearly visible event in May 2012.

ESA's *Venus Express* spacecraft was launched in November 2005. Following a short cruise phase, the spacecraft entered orbit around Venus in April 2006 where it remained until the end of its lifetime around December 2014. *Venus Express* carried instruments for the study of the Venusian atmosphere and magnetosphere (Titov et al., 2006). Data from VEX's magnetometer (MAG; Zhang et al., 2006) have been used in this study, obtained from ESA's Planetary Science Archive (PSA; sciops.esa.int/index.php?project=PSA&page=vex). A time-averaged data resolution of 4 seconds has been used; figures displaying VEX data in this work show data at this resolution.

The heliocentric distance of VEX is shown by the red line in Figure 1, and data gaps by (light) grey regions in the top half of the figure. Data is available for the entire period of time that VEX spent at Venus, although none is available from the cruise phase. The VEX ICME catalogue presented here spans the period from May 2006 to December 2013.

[*Figure 1*]

## 3 Identification criteria

Three criteria were used to identify ICMEs in magnetometer data from MESSENGER and *Venus Express*, namely (i) relatively smooth, monotonic rotations of the magnetic field direction, coinciding with (ii) an enhanced magnetic field strength relative to that of the ambient solar wind field, which (iii) last for a period no less than 4 hours (~ 0.17 days). By adopting these criteria, we have sought to identify ICMEs that are possibly magnetic clouds also. The first two criteria are the magnetic field signatures of a magnetic cloud as originally described by Burlaga et al. (1981). The third criterion sets a different time threshold to that of Burlaga et al., who specify a duration of ~1 day for the field enhancement and rotation. A 1-day threshold is suitable for identifying magnetic clouds at 1 AU but clouds observed closer to the Sun may be missed with such a threshold, since magnetic clouds are often smaller closer to the Sun (and hence spacecraft intersect them for shorter durations). Given that MESSENGER and *Venus Express* spent much of their time at heliocentric distances significantly under 1 AU, the duration threshold has been reduced. The choice of a 4-hour minimum duration is somewhat arbitrary. The purpose of this criterion is to ensure that all large-scale flux ropes likely to be associated with CMEs are included in the catalogue, whilst ensuring that the smaller-scale, non CME-related flux ropes which pervade the solar wind are excluded (Moldwin et al., 2000; Janvier et al., 2014). Flux rope boundaries have been located at discontinuous features in the field components, between which the field magnitude is enhanced and the field direction rotates in a way consistent with the identification criteria. The adoption of flux rope signatures as an ICME identifier is also an effective way to exclude field enhancements associated with stream and co-rotating interaction regions (SIRs and CIRs), which do not show such a field configuration. No strict value was placed on the field enhancement factor relative to the solar wind in order to meet the second criterion.

Figure 2 shows an example of an ICME observed by MESSENGER in January 2009, during the spacecraft's cruise phase. The boundaries of the ICME's flux rope are indicated with vertical dotted lines. The first panel displays the magnetic field magnitude. At its peak within the flux rope, the field magnitude was ~3.5 times the value it held in the ambient solar wind, clearly satisfying the second identification criterion.

The second panel shows the three components of the magnetic field, ***B***, in RTN co-ordinates. The *R* direction in this spacecraft-centred co-ordinate system points radially away from the Sun, *T* points in the direction of the vector product of *R* with the solar rotation axis, and *N* completes the right-handed set. The third panel shows the latitude angle of the B-field vector relative to the *R-T* plane, $\theta_{RTN} = \sin^{-1}(B_N/B)$; $\theta_{RTN} = 90°$ indicates that the vector points in the +*N* direction, $\theta_{RTN} = -90°$ in the –*N* direction. The fourth panel shows the angle between the projection of ***B*** onto the *R-T* plane and the *R* direction, $\varphi_{RTN}$, such that

$$\varphi_{RTN} = \begin{cases} \tan^{-1}(B_T/B_R), & \text{if } B_R > 0 \\ \tan^{-1}(B_T/B_R) + 180°, & \text{if } B_R < 0 \text{ and } B_T \geq 0 \\ \tan^{-1}(B_T/B_R) - 180°, & \text{if } B_R < 0 \text{ and } B_T < 0 \end{cases}$$

$\varphi_{RTN}$ has been mapped to angles between 0 and 360° in Figure 2. Thus, for a B-field vector laying in the *R-T* plane (*i.e.*, when $\theta_{RTN} = 0$), $\varphi_{RTN}$ values of 0, 90°, 180° and 270° correspond to the +*R*, +*T*, –*R* and –*T* directions, respectively.

Field rotations are easily discerned by plotting $\theta_{RTN}$ and $\varphi_{RTN}$: they show a relatively smooth and monotonic rotation of the B-field direction within the flux rope in Figure 2 (satisfying the first identification criterion), over the course of 1.25 days (satisfying the third criterion). The discontinuities in the field components near the middle of the rope around day of year (DOY) 20.6 are consistent with the formation of a current sheet, suggesting that the flux rope cross section may have been flattened and non-circular (Owens, 2009). Note also the relatively gradual rise in the B-field strength ahead of the rope in Figure 2, and the absence of any significant shock or discontinuity. Many ICMEs do not drive shocks at times close to solar minimum (such as when this event was observed), and so a shock-driving condition for identifying ICMEs has not been imposed.

[*Figure 2*]

Figure 3 shows an ICME that did drive a shock.  This event was observed by *Venus Express* in September 2006.  The shock ahead of the flux rope (at ~DOY 253.6) is marked by a dashed line, and the flux rope boundaries by two dotted lines as in Figure 2.  The magnetic field components in the second panel are displayed in Venus Solar Orbital (VSO) co-ordinates: *x* points from Venus towards the Sun, *y* from Venus in the direction opposite to the planet's orbital motion, and *z* completes the right-handed set.  $\theta_{VSO}$ and $\varphi_{VSO}$, shown in the third and fourth panels respectively, are equivalent to $\theta_{RTN}$ and $\varphi_{RTN}$ in the RTN system.  $\theta_{VSO} = 90°$ indicates that **B** points in the +z direction, $\theta_{VSO} = -90°$ in the –z direction; when in the *x-y* plane, $\varphi_{VSO}$ values of 0, 90°, 180° and 270° indicate a vector pointing in the +*x*, +*y*, –*x* and –*y* directions, respectively.  Data obtained by the spacecraft when it was close to the induced magnetosphere of Venus have been excluded from the plot; these intervals, which have been identified manually, are shaded grey.  Bow shock crossings are indicated with dash-dotted lines.

The three ICME identification criteria were clearly met by this event: the field magnitude at the rope centre was enhanced by a factor of ~7 relative to the solar wind, and the field rotated relatively smoothly for ~0.8 days.  Data obtained when the spacecraft was within the planet's magnetosheath – *i.e.*, the interval either side of the grey-shaded regions, up to the bow shock crossings – have been retained since the field rotations of the flux rope were still apparent in these intervals.  Note that the enhanced field magnitude in these regions is a result of the bow shock crossings, and is not an intrinsic feature of the flux rope field.

[*Figure 3*]

Figure 4 shows another ICME observed by *Venus Express*, in October 2011.  Boundaries within the data are marked in the same way as for Figure 3.  Like the previous examples, this event clearly met the criteria for classification as an ICME, despite the flux rope boundaries being partly obscured in this case.  VEX was close to Venus's induced magnetosphere during the arrival of the flux rope and so the leading edge was not directly observed.  The trailing edge was observed at DOY 209.4, soon after the spacecraft made an outbound bow shock crossing.  There is a large kink in $\varphi_{RTN}$ near the rope centre, a result of significant discontinuities in the three field components in this region.  Again, this points to the presence of a current sheet and non-circular flux rope cross section.

[*Figure 4*]

## 4   Analysis

Table 1 lists details of the ICMEs found using magnetic field data from MESSENGER and *Venus Express*.  Events in the tables are given a 3-part identification number, based on the observing spacecraft ('MES' or 'VEX'), the year of the observation, and the chronology of the event relative to other events observed that year.  For example, event MES201103 was the third ICME observed by MESSENGER in 2011.  Where two events appear in close temporal proximity to each other in the data, each rope ID is given an 'a', 'b' … etc. suffix.

For each ICME, the times at which the spacecraft observed the leading and trailing edges of the flux rope have been given.  Trailing edges are not always straightforward to identify, particularly when only magnetic field data is available to make the identification: for some ambiguous cases, alternative trailing edge locations have therefore been suggested.  Where a boundary is obscured by a data gap or (more likely) by the spacecraft being out of the solar wind, a range of times within which the boundary would have been observed is given.  The duration of the flux rope interval is also indicated, as well as the maximum field strength reached within the rope.  Where the flux rope drove a shock or discontinuity, the time at which the shock or discontinuity was observed is given.  Values in parentheses are less certain. Table 1 is also included as part of the electronic supplementary material that accompanies the version of this work published by *Solar Physics*.

Each event observed by the spacecraft has been given a quality rating value of '1', '2', or '3', denoting events of high, intermediate and low quality, respectively. A quality 1 event is characterised by particularly smooth and low variance rotation, sharply defined flux rope boundaries, and few or no data gaps in its data time series. A quality 3 event is characterised by field rotations with higher than ideal variance (which are, nonetheless, still consistent with a flux rope structure), a weak field magnitude enhancement, or data gaps and obscured rope boundaries. Quality 2 events have some characteristics of both quality 1 and 3 events. Figures 2 and 3 show examples of quality 1 events, and Figure 4 presents a quality 2 event. The rating encompasses both the quality of the observations made by the spacecraft (*e.g.*, whether data gaps were present or rope boundaries observed) and the intrinsic quality of the ICME itself (*e.g.*, the degree of smoothness in the rotation, or the field enhancement factor relative to the solar wind.) There is inevitably a degree of subjectivity in rating events in this manner.

A small number of events have been included in the catalogues that only strictly meet two of the three identification criteria. These events are indicated by an asterisk (*) next to their ID numbers in Table 1. For example, event MES201201 (previously studied by Rollett et al., 2014) showed a very strong field magnitude enhancement over 0.43 days, with the relatively smooth rise and fall in the field magnitude often seen in magnetic clouds, but the field direction showed little rotation. In this case, the spacecraft could have intersected a leg of the ICME connecting the flux rope back to the Sun. Likewise, event VEX201007b (previously studied by Möstl et al., 2012) displayed a large field magnitude enhancement with flux rope-like rotations, but its duration was only 0.14 days. This may have been a particularly fast ICME. Such events have been included since they are likely to be ICMEs despite failing to meet one of the criteria; the criteria that are met by these events are met unambiguously.

| Year/Event # | $r_H$, AU | Discontinuity, DOY | Rope leading edge, DOY | Rope trailing edge, DOY | Rope trailing edge (*alt.*), DOY | Duration, days | $B_{max}$, nT | Quality | Rope configuration | Handedness |
|---|---|---|---|---|---|---|---|---|---|---|
| | | | | MESSENGER | | | | | | |
| MES… | | | | | | | | | | |
| 200501 | 0.925 | 128.513 | 128.668 | 129.126 | 129.939 | 0.458 | 31 | 1 | NWS | L |
| 200502 | 0.928 | 134.951 | 135.350 | 136.058 | 137.514 | 0.708 | 38 | 1 | ENW | L |
| 200701 | 0.587 | - | 124.873 | 126.040 | 126.248 | 1.167 | 23 | 2 | NWS | L |
| 200702 | 0.669 | - | 144.948 | 145.246 | 145.719 | 0.298 | 20 | 2 | SWN | R |
| 200703a | 0.743 | - | 167.214 | 167.923 | - | 0.709 | 10 | 1 | SWN | R |
| 200703b | 0.744 | - | 167.944 | 168.377 | - | 0.433 | 17 | 1 | (WNE) | R |
| 200801 | 0.692 | - | 97.728 | 98.051 | - | 0.323 | 26 | 1 | WSE | L |
| 200802 | 0.526 | (260.074) | 260.302 | 260.809 | 260.590 | 0.507 | 31 | 2 | NES | R |
| 200801 | 0.723 | - | 364.865 | (365.207) | 365.322 to 365.403 | 0.342 | 11 | 2 | NWS | L |
| 200901 | 0.444 | - | 20.253 | 20.996 | 21.030 | 0.743 | 38 | 1 | NWS | L |
| 200902 | 0.327 | - | 46.183 | 46.592 | 46.794 | 0.410 | 47 | 3 | - | - |
| 200903 | 0.371 | - | 51.817 | 52.014 | - | 0.197 | 40 | 3 | NWS | L |
| 200904 | 0.625 | - | (223.688) | 224.252 | - | 0.564 | 24 | 2 | NES | R |
| 200905 | 0.561 | - | 240.057 | 240.650 | 240.890 | 0.593 | 20 | 1 | NWS | L |
| 200906 | 0.343 | - | 266.580 | 267.287 | - | 0.707 | 37 | 2 | NES | R |
| 200907 | 0.439 | - | 360.690 | 361.765 | - | 1.075 | 24 | 2 | (ENW) | L |
| 201001 | 0.552 | 81.156 | 81.368 | 81.920 | 81.982 | 0.552 | 32 | 1 | NWS | L |
| 201002 | 0.562 | - | 170.445 | 170.993 | - | 0.548 | 31 | 2 | - | - |
| 201003 | 0.378 | - | 242.704 | 243.526 | - | 0.822 | 35 | 2 | NES | R |
| 201004 | 0.462 | 309.490 | 309.703 | 310.547 | - | 0.844 | 55 | 1 | NES | R |
| 201005 | 0.337 | - | 343.075 | 343.601 | 343.507 | 0.526 | 65 | 1 | - | - |
| 201006 | 0.365 | 347.178 | 347.489 | 347.698 | - | 0.209 | 54 | 2 | - | - |
| 201101 | 0.525 | - | 5.345 | 6.165 | - | 0.820 | 27 | 2 | NES | R |
| 201102 | 0.511 | - | 43.229 | 43.734 | - | 0.505 | 47 | 1 | NES | R |
| 201103 | 0.331 | - | 68.089 | 68.520 | 68.567 | 0.432 | 91 | 3 | NWS | L |
| 201104 | 0.407 | 139.493 | 139.698 | 140.2 to 140.443 | - | 0.624 | ? | 3 | - | - |
| 201105 | 0.322 | (156.147) | 156.188 | 156.353 to 156.534 | - | 0.256 | 172 | 2 | - | - |
| 201106 | 0.460 | 288.352 | 288.470 | 289.266 | - | 0.796 | 90 | 1 | WSE | L |
| 201107 | 0.439 | 308.631 | 309.030 | 309.712 | 309.571 | 0.682 | 42 | 1 | NWS | L |
| 201108 | 0.396 | - | 318.272 to 318.445 | 318.734 to 318.957 | 319.000 | 0.487 | 50 | 3 | NWS | L |
| 201109 | 0.374 | - | 321.763 to 321.954 | (322.600) | - | 0.742 | 56 | 3 | - | - |
| 201110 | 0.345 | 327.211 | 327.263 to 327.452 | 327.626 to 327.945 | - | 0.428 | 100 | 3 | SEN | L |
| 201111 | 0.424 | 364.686 | 364.884 | 365.389 | - | 0.505 | 65 | 2 | - | - |
| 201201* | 0.320 | 67.211 | 67.257 | (67.622 to 67.751) | - | 0.430 | 130 | 2 | - | - |
| 201202 | 0.464 | 100.217 | 100.373 | 101.275 | 101.242 | 0.902 | 52 | 2 | SEN | L |
| 201203 | 0.310 | - | 146.349 | 147.475 | - | 1.126 | 101 | 2 | SWN | R |

*Venus Express*

VEX...

| | | | | | | | | | | |
|---|---|---|---|---|---|---|---|---|---|---|
| 200601 | 0.724 | - | 186.196 | 186.486 | - | 0.290 | 25 | 3 | (NES) | R |
| 200602 | 0.723 | 198.184 | 198.261 | 198.756 | 199.438 | 0.495 | 22 | 2 | NWS | L |
| 200603 | 0.718 | 253.591 | 253.762 | 254.547 | 254.695 | 0.785 | 36 | 1 | NWS | L |
| 200701 | 0.725 | - | 44.611 | 45.398 | 45.106 | 0.787 | 19 | 2 | WNE | R |
| 200702* | 0.719 | - | (117.370) | 117.674 | - | 0.304 | 13 | 3 | (NWS) | L |
| 200703 | 0.719 | - | 126.202 to 126.269 | 126.870 | - | 0.635 | 13 | 2 | NWS | L |
| 200704 | 0.721 | - | 145.150 | 145.833 | - | 0.683 | 16 | 2 | SWN | R |
| 200705a | 0.724 | - | 167.094 | 167.930 | - | 0.836 | 11 | 2 | SWN | R |
| 200705b | 0.724 | - | 167.980 | 168.446 | - | 0.466 | 17 | 2 | NES | R |
| 200706 | 0.722 | - | 285.725 | 285.899 | - | 0.174 | 21 | 1 | NWS | L |
| 200707* | 0.719 | - | 321.306 | 321.792 | - | 0.486 | 20 | 2 | - | - |
| 200708 | 0.719 | - | 341.756 | 343.614 | 342.626 | 1.858 | 19 | 2 | NWS | L |
| 200801* | 0.723 | - | 364.865 | (365.207) | 365.322 to 365.403 | 0.342 | 11 | 2 | NWS | L |
| 200901 | 0.721 | - | 17.609 | 17.951 | - | 0.343 | 13 | 2 | NWS | L |
| 200902 | 0.725 | (118.902) | 118.996 to 119.331 | 120.214 | - | 1.051 | 14 | 3 | NES | R |
| 200903 | 0.727 | - | 136.901 | 137.682 | 137.800 | 0.781 | 14 | 2 | NWS | L |
| 200904 | 0.728 | - | 153.777 | 154.514 | - | 0.737 | 12 | 2 | WNE | R |
| 200905 | 0.728 | - | 158.249 | 158.607 | - | 0.358 | 12 | 1 | NWS | L |
| 200906 | 0.728 | - | 175.240 | 175.490 | - | 0.250 | 13 | 2 | (NWS) | L |
| 200907 | 0.727 | - | 191.443 | 192.267 | 192.360 | 0.824 | 16 | 2 | NWS | L |
| 200908 | 0.723 | - | 221.436 | 221.803 | - | 0.367 | 17 | 2 | NWS | L |
| 200909 | 0.720 | - | 248.836 | 249.015 to 249.099 | - | 0.221 | 21 | 2 | NES | R |
| 200910 | 0.719 | - | 290.240 | 290.963 | 290.750 | 0.723 | 10 | 1 | (SEN) | L |
| 200911 | 0.722 | - | 325.410 | 325.666 | 326.716 | 0.256 | 13 | 1 | NWS | L |
| 201001 | 0.725 | - | (62.652) | 63.150 | - | 0.498 | 29 | 2 | NWS | L |
| 201002 | 0.724 | - | 76.388 | 77.510 | - | 1.122 | 21 | 1 | NES | R |
| 201003 | 0.719 | 157.569 | 157.608 | 158.031 | - | 0.423 | 22 | 1 | (ESW) | R |
| 201004 | 0.720 | - | 166.973 | 168.072 | - | 1.099 | 14 | 2 | - | - |
| 201005 | 0.721 | - | 173.620 | 174.770 | - | 1.150 | 21 | 2 | - | - |
| 201006 | 0.722 | - | 180.360 | 180.959 | - | 0.599 | 14 | 2 | - | - |
| 201007a | 0.726 | 213.612 | 213.966 | 214.247 | - | 0.281 | 27 | 2 | NWS | L |
| 201007b* | 0.726 | 214.479 | 214.529 | 214.670 | - | 0.141 | 60 | 2 | WSE | L |
| 201007c | 0.726 | - | 214.921 | 215.604 | - | 0.683 | 28 | 1 | - | - |
| 201008 | 0.727 | 222.159 | 222.522 | 222.885 | - | 0.363 | 27 | 1 | WSE | L |
| 201009* | 0.727 | - | 224.930 | (225.222) | - | 0.292 | 9 | 3 | - | - |
| 201010* | 0.728 | - | 251.461 | 251.773 | - | 0.312 | 11 | 2 | - | - |
| 201011 | 0.728 | 253.672 | 253.915 | 254.714 | - | 0.799 | 20 | 2 | WSE | L |
| 201012 | 0.719 | - | 344.845 | 345.923 | 345.708 | 1.078 | 19 | 2 | - | - |
| 201101 | 0.727 | (77.866) | 78.138 to 78.248 | 78.496 | - | 0.302 | 28 | 3 | NWS | L |
| 201102 | 0.727 | 81.369 | 81.728 | 82.764 | - | 1.037 | 18 | 2 | - | - |
| 201103 | 0.728 | - | 95.717 | 96.627 | - | 0.910 | 20 | 3 | NES | R |
| 201104 | 0.728 | - | 101.461 | 101.731 | - | 0.270 | 15 | 1 | NWS | L |
| 201105 | 0.728 | - | 107.319 | 107.762 | - | 0.443 | 16 | 2 | - | - |
| 201106a | 0.728 | - | 110.286 | 110.917 | 111.074 to 111.226 | 0.631 | 16 | 3 | NWS | L |
| 201106b | 0.728 | - | 111.472 | 112.477 | - | 1.005 | 17 | 2 | NES | R |
| 201107 | 0.727 | 139.874 | (140.161) | 140.759 | - | 0.598 | 27 | 2 | NES | R |
| 201108 | 0.724 | 156.562 | 156.634 | 156.938 | - | 0.304 | 59 | 1 | - | - |
| 201109 | 0.721 | - | 182.573 | 183.376 | - | 0.803 | 18 | 2 | (NWS) | L |
| 201110 | 0.723 | - | 274.050 | 274.613 | - | 0.563 | 16 | 2 | SWN | R |
| 201111 | 0.725 | 289.035 | 289.254 to 289.326 | 290.402 | 290.24 to 290.383 | 1.112 | 43 | 2 | WSE | L |
| 201112a | 0.727 | 309.154 | 309.289 to 309.461 | 309.659 | 310.009 | 0.284 | 53 | 2 | NES | R |
| 201112b | 0.727 | - | 310.077 | 311.132 | - | 1.055 | 14 | 2 | NWS | L |
| 201113 | 0.728 | (323.158) | 323.319 to 323.451 | 324.138 | - | 0.753 | 40 | 2 | (NES) | R |
| 201114a | 0.727 | 359.527 | 359.652 | 360.032 | - | 0.380 | 20 | 2 | NES | R |
| 201114b | 0.727 | - | (360.118) | 360.943 | 360.699 | 0.581 | 31 | 3 | ESW | R |
| 201114c | 0.727 | - | 361.097 | 361.654 | - | 0.557 | 20 | 2 | NES | R |
| 201201 | 0.724 | - | 16.931 to 17.253 | 18.091 | - | 0.999 | 42 | 3 | SWN | R |
| 201202 | 0.722 | - | 32.880 | 33.559 | - | 0.679 | 23 | 3 | - | - |
| 201203 | 0.721 | - | 43.2589 to 43.3186 | 43.564 | 43.574 | 0.275 | 22 | 3 | - | - |
| 201204 | 0.719 | 61.568 | 61.894 | 62.556 | 62.769 | 0.663 | 23 | 3 | SWN | R |
| 201205* | 0.719 | 67.560 | 67.843 | 68.488 | 68.976 | 0.645 | 22 | 2 | - | - |
| 201206 | 0.722 | (123.027) | 123.634 | 125.064 | - | 1.430 | 19 | 2 | NES | R |
| 201207 | 0.727 | - | 167.810 | 168.353 | 168.367 | 0.543 | 88 | 1 | NES | R |
| 201208 | 0.728 | - | 189.847 | 190.043 to 190.172 | - | 0.261 | 30 | 3 | NWS | L |
| 201209* | 0.728 | 200.693 | 200.994 to 201.235 | 201.697 | - | 0.582 | 15 | 2 | - | - |
| 201210 | 0.728 | - | 211.458 | 211.749 | - | 0.291 | 40 | 1 | (NES) | R |
| 201211 | 0.722 | 257.197 | 257.489 | 258.137 | 258.319 | 0.648 | 21 | 2 | NES | R |
| 201212 | 0.719 | 315.652 | 315.914 | 316.080 to 316.140 | 316.199 | 0.196 | 24 | 3 | SWN | R |
| 201213 | 0.719 | 318.450 | 318.730 | 319.269 | - | 0.539 | 31 | 1 | WNE | R |
| 201214 | 0.720 | (330.043) | 330.220 | 330.472 | 330.647 | 0.252 | 16 | 2 | NES | R |
| 201215 | 0.722 | - | 352.513 | 353.155 to 353.265 | - | 0.697 | 26 | 3 | - | - |
| 201301 | 0.725 | 8.391 | 8.642 | 9.825 | - | 1.183 | 29 | 1 | WNE | R |
| 201302 | 0.728 | - | 33.094 | 33.869 | - | 0.775 | 16 | 1 | NES | R |
| 201303* | 0.728 | - | 48.455 | 48.869 | - | 0.413 | 13 | 2 | SWN | R |
| 201304 | 0.728 | - | 52.669 | (52.861) | - | 0.192 | 20 | 3 | ENW | L |
| 201305 | 0.728 | 65.557 | 65.983 | 66.485 | 67.193 to 67.317 | 0.502 | 28 | 3 | - | - |
| 201306 | 0.727 | - | 72.862 | 73.303 | - | 0.441 | 23 | 1 | ENW | L |
| 201307 | 0.722 | 117.610 | 118.073 | (119.579) | - | 1.506 | 19 | 2 | SEN | L |
| 201308* | 0.721 | 201.421 | 201.882 | 202.741 | - | 0.859 | 13 | 3 | - | - |
| 201309 | 0.728 | - | (261.487) | 261.939 | - | 0.452 | 19 | 2 | NES | R |
| 201310 | 0.728 | 278.179 | 278.523 | (279.624) | 279.206 | 1.101 | 19 | 2 | - | - |
| 201311 | 0.727 | (300.468) | 301.076 | 302.212 | - | 1.136 | 44 | 1 | (SEN) | L |
| 201312 | 0.723 | 334.188 | 334.583 | 335.667 | - | 1.084 | 24 | 1 | NES | R |
| 201313* | 0.721 | 348.130 | 348.606 | 348.986 | - | 0.380 | 16 | 2 | - | - |

**Table 1.** The MESSENGER and *Venus Express* ICME catalogue. Parenthesised table entries indicate uncertainty in the stated values.

## 4.1 Event statistics

A total of 35 ICMEs found in the MESSENGER data and a total of 84 ICMEs found in the *Venus Express* data are listed in Table 1. The number of ICMEs encountered by the two spacecraft in each year is given in Table 2, along with the mean heliocentric distances at which the ICMEs were observed, $\langle r_H \rangle$, and the mean maximum B-field strengths within the ICMEs' flux ropes, $\langle B_{max} \rangle$. MESSENGER encountered events across the full range of heliocentric distances it covered during its cruise phase and subsequent insertion at Mercury; all of the events encountered by *Venus Express* were observed at ~0.72 AU, the heliocentric distance of Venus. The MESSENGER catalogue contains fewer events than the *Venus Express* catalogue, probably a reflection of the shorter timespan it covers compared to the VEX catalogue (see Section 2). Italicised event numbers in Table 2 indicate that data for that year is incomplete (*i.e.*, 2005 and 2007 for MESSENGER, and 2006 for VEX), or that the data has only been partly analysed (*i.e.*, 2012 for MESSENGER).

There is some limited evidence of a solar cycle dependence in the figures presented in Table 2. Comparing those years with largely complete datasets, it can be seen that fewer events were observed per year around the deep solar minimum of 2008 than in the declining phase of the previous cycle (Cycle 23) and the ascending phase of the current cycle (Cycle 24). There is also a general increase in $\langle B_{max} \rangle$ from 2008 as solar maximum is approached. Note the increasing trend in $\langle B_{max} \rangle$ observed by MESSENGER was partly an effect of the spacecraft's declining heliocentric distance during this period – ICME flux ropes closer to the Sun tend to have larger B-field magnitudes that subsequently fall as the ropes expand during propagation.

| Year | MESSENGER | | | *Venus Express* | | |
|---|---|---|---|---|---|---|
| | # of events | $\langle r_H \rangle$ (AU) | $\langle B_{max} \rangle$ (nT) | # of events | $\langle r_H \rangle$ (AU) | $\langle B_{max} \rangle$ (nT) |
| 2005 | *2* | 0.93 | 35 | - | | |
| 2006 | - | | | *3* | 0.72 | 28 |
| 2007 | *4* | 0.69 | 18 | 9 | 0.72 | 18 |
| 2008 | 2 | 0.61 | 29 | 1 | 0.72 | 11 |
| 2009 | 7 | 0.44 | 33 | 11 | 0.72 | 14 |
| 2010 | 6 | 0.44 | 35 | 14 | 0.72 | 23 |
| 2011 | 11 | 0.41 | 74 | 18 | 0.73 | 26 |
| 2012 | *3* | 0.36 | 94 | 15 | 0.72 | 29 |
| 2013 | - | | | 13 | 0.73 | 22 |

**Table 2.** The ICME catalogue in summary. Italicized numbers indicate that that year's dataset for the spacecraft in question is incomplete or has not been analysed in full.

## 4.2 Flux rope orientations

A simple way to categorise ICME flux ropes is by their approximate configuration relative to the SE plane. This configuration may be described in terms of the rotation of the magnetic field vector as the rope traverses the spacecraft. Using the three-part notation adopted by Mulligan et al. (1998), which indicates the field directions of the successively observed leading edge, centre and trailing edge of the rope, the configurations of the ropes observed by MESSENGER and *Venus Express* have been determined; these configurations are listed in Table 1. The field directions are given in terms of the cardinal directions from the perspective of an observer looking towards the Sun, relative to the SE plane. For example, a rope designated 'NWS' has a predominantly northward leading edge field (*i.e.*, parallel to the solar rotation axis), a westward field at the centre (*i.e.*, pointing towards the western limb of the Sun) and a southward field at the trailing edge (*i.e.*, anti-parallel to the solar rotation axis). These cardinal directions may be approximately related to the RTN and VSO field angles defined previously: N ≡ $\theta_{RTN}$ ~ $\theta_{VSO}$ ~ 90°; S ≡ $\theta_{RTN}$ ~ $\theta_{VSO}$ ~ -90°; E ≡ $[\theta, \varphi]_{RTN}$ ~ [0°, 270°] and $[\theta, \varphi]_{VSO}$ ~ [0°, 90°]; and W ≡ $[\theta, \varphi]_{RTN}$ ~ [0°, 90°] and $[\theta, \varphi]_{VSO}$ ~ [0°, 270°]. These approximations may reasonably be made because the orbital planes of MESSENGER and *Venus Express* never deviated significantly (at least on heliospheric scales) from the SE plane. Figures 2 and 3 show examples of NWS flux rope configurations, and Figure 4 an example of a WSE configuration.

Ropes which were poorly observed or with complex rotations that do not conform to a simple flux rope picture (around 25% of catalogued events) have not been categorised in this way. Some of these uncategorised events showed strongly radial fields with little rotation (*e.g.*, event MES201201); such events are sometimes described as 'magnetic cloud-like' (*e.g.*, Wu and Lepping, 2007). Others may have been 'complex ejecta', *i.e.*, structures consisting of multiple flux ropes, or ICMEs that has been distorted through interactions with other ICMEs or the solar wind.

### 4.2.1 Northward *vs.* southward leading edges

The numbers of each configuration type observed by both spacecraft are summarised in Table 3. It has been found that, across both spacecraft, ropes with northward leading edges (NES or NWS) were observed around 4 times more often than ropes with southward leading edges (SEN or SWN). This finding is in agreement with the previously established rule that NES/NWS (SEN/SWN) magnetic clouds predominate from the late declining phase of an odd (even) numbered solar cycle to the rising phase of the next cycle (Bothmer and Schwenn, 1998; Mulligan et al., 1998; Li and Luhmann, 2004; Huttunen et al., 2005): all of the ICME flux ropes listed in this work were observed in either the declining phase of Cycle 23 or rising phase of Cycle 24. An analogous result was, for example, found by Bothmer and Schwenn in their study of magnetic clouds observed by *Helios* during the declining phase of Cycle 20 and rising phase of Cycle 21, when SEN/SWN clouds were around three times more common than NES/NWS clouds.

The predominance of NES/NWS over SEN/SWN ropes may be a result of the global configuration of the solar magnetic field (Mulligan et al., 1998). During the period of time studied, the solar northern hemisphere largely consisted of inward-directed magnetic field and the southern hemisphere of outward-directed field. If most flux ropes are thought to emerge from the streamer belt around the solar equator, then such a solar field topology would lead to the greater instances of NES and NWS ropes that were observed. Li et al. (2011) highlight some of the nuances in this interpretation.

### 4.2.2 High *vs.* low inclination to the SE plane

An imbalance also exists between the number of flux ropes observed with an east/west axis orientation (NES, SEN, NWS and SWN) and those with a north/south axis orientation (ENW, WNE, ESW and WSE): the former were observed about four times more frequently than the latter (see Table 3). This imbalance may in part be due to the relative likelihood of observation of each flux rope orientation. East/west-oriented ropes, with their axes at low inclinations to the SE plane, intersect a larger area of the SE plane close to which the observing spacecraft are found, and so are more likely to be observed. Conversely, north/south oriented ropes intersect a smaller area of the SE plane, and so are less likely to be observed. This interpretation assumes that ICME flux ropes are wider in the direction parallel to the flux rope axis than in the direction transverse to the rope axis.

Some have suggested that the east/west direction is a preferential direction for flux ropes observed in the heliosphere. Yurchyshyn (2008) and Isavnin et al. (2014), amongst others, suggest that ICME flux ropes can rotate during propagation to align with the local direction of the heliospheric current sheet, which often lies close to the SE plane.

### 4.2.3 Left *vs.* right-handedness

Each flux rope configuration is associated with an intrinsic handedness. Handedness indicates the sense of field-line winding around the axis of the rope. NWS, SEN, ENW and WSE ropes are left-handed, and NES, SWN, ESW and WNE ropes are right-handed. It has been found that left-handed and right-handed flux ropes were observed in almost equal numbers (46 and 44, respectively) by MESSENGER and *Venus Express*. Given the solar-hemispherical ordering of ICME flux rope handedness (Rust, 1994), this result is to be expected.

|  | MESSENGER | *Venus Express* | Total |
|---|---|---|---|
| A. Northward leading edge (NES, NWS) | 16 | 42 | 58 |
| B. Southward leading edge (SEN, SWN) | 5 | 10 | 15 |
| C. High inclination (WNE, WSE, ENW, ESW) | 5 | 12 | 17 |
| D. Left-handed (NWS, SEN, ENW, WSE) | 15 | 31 | 46 |
| E. Right-handed (NES, SWN, ESW, WNE) | 11 | 33 | 44 |
| F. Uncategorised (-) | 9 | 20 | 29 |
| A : B | 3.2 : 1 | 4.2 : 1 | 3.9 : 1 |
| [A and B] : C | 4.2 : 1 | 4.3 : 1 | 4.3 : 1 |
| D : E | 1.4 : 1 | 0.9 : 1 | 1.0 : 1 |

**Table 3.** The distribution of ICME flux rope configurations observed by MESSENGER and *Venus Express*.

## 4.3 Multipoint observations

For each of the MESSENGER events listed in Table 1, *in situ* data from *Venus Express*, STEREO-A (STA), STEREO-B (STB) and ACE have been examined for signatures of the same event. Likewise, *in situ* data from MESSENGER and the three 1 AU spacecraft have been examined for signatures of the *Venus Express* events. STEREO data have been obtained from the Space Physics Center of the UCLA Institute of Geophysics and Planetary Physics (aten.igpp.ucla.edu), and ACE data have been obtained from the ACE Science Center (www.srl.caltech.edu/ACE/ASC/index.html).

Data at each spacecraft have been examined at time windows within which the ICMEs observed at MESSENGER or *Venus Express* would be expected to arrive. For example, an ICME observed by MESSENGER at Mercury would have been likely to arrive at Venus 1 – 2 days later, and ~3 days later at 1 AU. Rigid search windows were not imposed in order to allow for particularly fast or slow events.

An event-by-event analysis is detailed in the electronic supplementary material accompanying the version of this work published by *Solar Physics*. It lists the heliocentric inertial (HCI) longitudes of MES ($\Theta_{MES}$) and VEX ($\Theta_{VEX}$) when they observed each of the ICMEs listed in Table 1, the HCI longitude of the other spacecraft at the expected arrival times of each ICME ($\Theta_x$), the longitudinal separation between each spacecraft pair ($\Delta\Theta_x$), and the signature type observed at the second spacecraft. The longitudinal separations represent the angles within the SE plane subtended at the Sun by the spacecraft pairs. Latitudinal separations have been neglected.

### 4.3.1 Signature classification

Signatures within the search windows at the other spacecraft have been given one of the following three classifications:

- Flux rope signatures are present and likely to be associated with the ICME observed at MES or VEX (denoted 'YY' in the supplementary table)
- A shock or discontinuity with a significant $|B|$ enhancement (without any flux rope signatures) is present and likely to be associated with the ICME observed at MES or VEX (denoted 'Y' in the supplementary table)
- No *in situ* signatures of an ICME are present, or, if present, are unlikely to be associated with the ICME observed at MES or VEX given coronagraph observations (denoted 'N' in the supplementary table)

Likelihood of association was assessed by considering expected arrival windows (as described in the second paragraph of this subsection) alongside coronagraph observations from LASCO (on board the SOHO spacecraft at L1), COR-2A, and COR-2B (on board STEREO-A and STEREO-B, respectively). A 'YY' or 'Y' classification has only been deemed likely if a single CME was launched from the Sun within a broad launch-time window, and if the apparent propagation direction of that CME - roughly estimated, from the vantage points of all three coronagraphs simultaneously - was towards both of the spacecraft in question. Greater emphasis was placed on coronagraph observations when determining associations for widely-spaced spacecraft (> ~60°). A classification is parenthesised in the supplementary table if:

(i) the coronagraph observations are inconclusive (*e.g.*, if multiple CMEs are launched in similar directions around the same time); or
(ii) the arrival time of the signature deviates significantly from what is expected.

Flux ropes were identified at the other spacecraft by their magnetic field signatures in the same way as for the MESSENGER and *Venus Express* event cataloguing. The 1 AU spacecraft used in this analysis all carry instruments for analysing the solar wind plasma; where a signature was observed at 1 AU, the *in situ* proton speed there has been compared to the mean speed between the two spacecraft (calculated from arrival times) as a further test of likelihood that the signature was associated with a particular ICME. $|B|$ enhancements that appeared to be associated with SIRs or CIRs have been excluded when making a 'Y' classification. Entries in the supplementary table marked 'X' indicate the presence of a significant data gap at the time when a signature would have been expected to appear. Italicised 'YY' entries in the VEX list indicate that the ICME's flux rope was also observed by MES.

### 4.3.2 An example event

Figure 5 shows an example of an ICME observed by MESSENGER (event MES201004 in Table 1) that was subsequently observed by STEREO-B. Magnetic field and plasma data (including bulk flow speed, density and temperature) from the cloud at STEREO-B are shown in Figure 6. The enhanced B-field magnitude and smooth field rotation characteristic of a magnetic cloud are present at both spacecraft; a magnetic cloud identification in this case is corroborated by the low plasma-beta ($\leq 0.1$) and temperature observed at STEREO-B. The two spacecraft were close to radial alignment around the time of observation, from 5 to 8 November 2010, with a separation of 1.3° in HCI latitude, 7.1° in HCI longitude and 0.61 AU in heliocentric distance. The cloud's sheath arrived at MES at DOY 309.490 followed by the flux rope's leading edge at DOY 309.703. Approximately 58 hours later, at DOY 312.119, the flux rope leading edge arrived at STEREO-B: the mean speed of the leading edge between the two spacecraft was thus ~440 km s$^{-1}$, comparable to the measured *in situ* speed at STEREO-B. The profiles of the magnetic field components and direction angles observed within the rope were similar at the two spacecraft; the rope had an NES configuration at both spacecraft. There may have been some compression (or suppressed expansion) at the rear of the flux rope due to a faster-moving solar wind stream catching up with the cloud at STEREO-B; a solar wind compression region is evident immediately to the rear of the rope.

The most likely CME counterpart to this cloud appeared in the LASCO C2 field of view on DOY 307.525 (3 November 2010, 12:36 UT) at the eastern limb of the Sun as viewed from L1. The CDAW CME catalogue (cdaw.gsfc.nasa.gov/CME_list) gives an estimated launch time from the solar surface of DOY 307.420 (November 3 2010, 10:05 UT) for this CME. The CME appeared as a halo in both the STEREO COR-2A and COR-2B fields of view. Taken together, the three sets of coronagraph observations suggest that the CME propagation direction was towards MES and STEREO-B. The estimated launch time and sheath arrival time at MES give a mean propagation speed of ~415 km s$^{-1}$ between the Sun and MES, similar to the MES to STEREO-B speed.

There was an ICME-like |***B***| enhancement (without 'YY' flux rope field rotations) at VEX and ACE at ~ DOY 311.3 and 312.4, respectively. VEX was at an 87° separation and ACE at an 84° separation from MES around this time. Arrival times alone suggest a possible association with the MES cloud. However, there is a second, Earth-directed CME appearing in the COR-2A -2B fields of view approximately 7 hours before the other CME, which may account for the VEX and ACE signatures. The signatures at VEX and ACE have therefore been given an '(N)' classification – *i.e.*, some signatures are present that are unlikely to be associated with the ICME under consideration given the coronagraph observations, but coronagraph observations are not conclusive given that two CMEs were launched from the Sun at similar times in not dissimilar directions.

[*Figure 5*]

[*Figure 6*]

### 4.3.3 Statistical findings

Table 4 shows the number of each signature type observed, binned according to longitudinal separation in 15°-wide bins. The secondary signatures of all the ICMEs listed in Table 1 have been grouped together in this table. The numbers of each signature type are presented in the first three table rows; parenthesised values in these rows include the less certain signature associations.

The fractional distribution of each signature type within each bin is given in the bottom six rows, where *P* gives the distribution of the values without parentheses in rows 1 to 3, and *P´* the distribution of values with parentheses. Binomial standard errors are given for each fraction. Some columns do not sum to 1 due to rounding.

|  | Spacecraft separation angle in the SE plane | | | | | |
|---|---|---|---|---|---|---|
| Signature type | 0 - 15° | 15 - 30° | 30 - 45° | 45 - 60° | 60 - 75° | 75°+ |
| Flux rope ('YY') | 32 (33) | 19 (20) | 7 (7) | 5 (5) | 0 (2) | 0 (2) |
| Sheath ('Y') | 6 (6) | 8 (8) | 4 (5) | 3 (4) | 2 (3) | 3 (6) |
| No signature ('N') | 1 (2) | 12 (14) | 28 (28) | 35 (35) | 33 (33) | 215 (221) |
| $P$['YY'] | 0.82 ± 0.06 | 0.49 ± 0.08 | 0.18 ± 0.06 | 0.12 ± 0.05 | 0 | 0 |
| $P$['Y'] | 0.15 ± 0.06 | 0.21 ± 0.06 | 0.10 ± 0.05 | 0.07 ± 0.04 | 0.06 ± 0.04 | 0.01 ± 0.01 |
| $P$['N'] | 0.03 ± 0.03 | 0.31 ± 0.07 | 0.72 ± 0.07 | 0.81 ± 0.06 | 0.94 ± 0.04 | 0.99 ± 0.01 |
| $P´$['YY'] | 0.80 ± 0.06 | 0.48 ± 0.08 | 0.18 ± 0.06 | 0.11 ± 0.05 | 0.05 ± 0.04 | 0.01 ± 0.01 |
| $P´$['Y'] | 0.15 ± 0.06 | 0.19 ± 0.06 | 0.13 ± 0.05 | 0.09 ± 0.04 | 0.08 ± 0.04 | 0.03 ± 0.01 |
| $P´$['N'] | 0.05 ± 0.03 | 0.33 ± 0.07 | 0.70 ± 0.07 | 0.80 ± 0.06 | 0.87 ± 0.05 | 0.97 ± 0.01 |

**Table 4.** The numbers of each signature type observed at other spacecraft (MESSENGER or *Venus Express*, STEREO-A, STEREO-B, and ACE) associated with the ICMEs observed by MESSENGER and *Venus Express*, binned according to spacecraft separation.

Figure 7 shows stacked bar plots of the fractional distributions in Table 4. Each bar is effectively normalised in height to the total number of spacecraft pairings within the corresponding bin; this number is indicated at the top of each bar. Figure 7(a) comprises the values without parentheses in Table 4, and Figure 7(b) the less certain, parenthesised values. Errors associated with the flux rope observations are displayed.

The trends in this figure are clear. As the spacecraft's longitudinal separation increased, the likelihood of the second spacecraft observing either the ICME's flux rope (bar components coloured blue) or sheath (coloured red) steadily dropped. At small angular separations, the second spacecraft almost always saw signatures of the same ICME's flux rope or sheath (95 or 97% of cases for separations between zero and 15°); conversely, a widely separated second spacecraft rarely saw any signature of the same ICME (1 or 4% of cases for separations greater than 75°).

Cane et al. (1997) found qualitatively similar results in their study of ICME ejecta (including magnetic clouds) observed by *Helios 1*, *Helios 2* and IMP 8. They found few ejecta were observed by spacecraft pairs even at 40° separations or less, and hardly any at separations above 40°. We note, however, that they studied relatively few events, with spacecraft separations only up to 53°.

Bothmer and Schwenn (1998) present a comparable plot to that shown in Figure 7 (Fig. 12 in their work), based upon *Helios* and IMP/ISEE observations of magnetic clouds. Although they considered fewer events, their findings are broadly consistent with those presented here. Of the fifteen clouds they considered, eight were observed by two spacecraft at separations less than 40°, one at a separation between 50 and 60°, and none above 60°; six clouds were only observed by a single spacecraft, all where the spacecraft separation was above 50°.

Figure 7 also shows that, at smaller spacecraft separations, ICME observations at the second spacecraft were more likely to be of the flux rope than of the sheath only. At the widest separations, conversely, it became more likely that the second spacecraft would miss the flux rope and encounter the sheath only. This finding is consistent with the notion that regions of solar wind disturbed by ICME propagation span a greater longitudinal extent that the ICME driving the disturbance, and that the shock or sheath region of an ICME extends beyond the flanks of the ICME's flux rope or ejecta (*e.g.*, Borrini et al., 1982).

[*Figure 7*]

## 5 Conclusion and discussion

Magnetometer data from the MESSENGER and *Venus Express* spacecraft have been examined for the signatures of ICMEs. An enhanced B-field magnitude and smoothly rotating B-field direction consistent with a magnetic flux rope structure were the primary identification criteria. A total of 35 ICMEs observed by MESSENGER within a 5-year period and 84 ICMEs observed by *Venus Express* within a 7.5-year period have been identified. A number of these ICMEs are likely to be magnetic clouds.

The flux rope within each ICME has been classified according to the observed magnetic field direction relative to the SE plane at its leading edge, centre, and trailing edge. It has been found that:

(i) ICME flux ropes with northward leading edges were about four times more common than flux ropes with southward leading edges. This result is consistent with the findings of Bothmer and Schwenn (1998), Mulligan et al. (1998), Li and Luhmann (2004), and Huttunen et al. (2005); it may be an effect of the underlying, global configuration of the solar magnetic field (Mulligan et al., 1998) during the ascending phase of Cycle 24, when magnetic field lines predominantly pointed inwards at the Sun's northern hemisphere and outwards at the Sun's southern hemisphere;

(ii) ICME flux ropes with low inclinations to the SE plane were about four times more common than flux ropes at high inclinations. This may partly be due to an observational bias – ICME flux ropes at high inclinations are likely to intersect a relatively smaller area of the SE plane, and so are less likely to be observed;

(iii) Left-handed ICME flux ropes were observed as frequently as right-handed flux ropes.

Multipoint observations for each of the catalogued ICMEs have been identified and classified. Data from MESSENGER, *Venus Express*, STEREO-A, STEREO-B and ACE were analysed, in conjunction with coronagraph observations. The results of this analysis are summarised in Table 4 and Figure 7. It has been found that:

(iv) The percentage of cases where a pair of spacecraft observed the flux rope or sheath of the same ICME decreased steadily as the longitudinal separation of the spacecraft increased. Where two spacecraft were separated by 15° or less, both spacecraft observed the ICME's flux rope in 82% of cases. This percentage dropped to 49% for spacecraft separations of 15 to 30°, to 18% for separations of 30 to 45°, and to 12% for separations of 45 to 60°;

(v) As the separation of a spacecraft pair became greater, it became increasingly likely that the second spacecraft would observe the sheath of an ICME rather than the flux rope, in those cases where the second spacecraft encountered any signature of the event. This supports the idea that the sheath regions of ICMEs have greater longitudinal extents than the ejecta.

This analysis may be useful for those wishing to determine the likelihood that two spacecraft separated in longitude will observe the same ICME; based upon these results, for example, a spacecraft stationed at L5 – *i.e.*, at a 60° separation from the Sun-Earth line – would be expected to encounter any Earth-incident ICMEs in less than ~10% of cases. Results (iv) and (v) are consistent with those obtained from earlier comparable studies (*e.g.*, Cane et al., 1997; Bothmer and Schwenn, 1998). However, the relatively large number of spacecraft-pair combinations considered in this analysis, made possible by the relatively large number of spacecraft available, gives the results a greater statistical robustness than the equivalent results obtained from the *Helios*-era studies.

The validity of the multipoint analysis that has been performed depends on the validity of the associations that have been made between *in situ* signatures at different spacecraft. For smaller spacecraft separations, such associations are relatively straightforward to make; there is greater ambiguity in associating signatures at widely separated spacecraft. Heliospheric imaging, or a more sophisticated modelling of CME propagation directions from the coronagraph images, could be used to confirm or otherwise some of the signature associations that have been made, particularly in those cases where the spacecraft separation was wide (*e.g.*, $> 45°$).

We suggest that this work could be used as the basis for a number of further studies:

- The ICMEs listed in Table 1 could be analysed using flux rope fitting techniques (*e.g.*, Lundquist fitting; Burlaga, 1988) to give more precise axis orientations. The evolution of axis orientation with heliocentric distance could be determined (*e.g.*, Leitner et al., 2007). However, some of the events listed (particularly those with obscured boundaries) would not be suited to such fitting;
- The event-by-event multipoint analysis presented in the supplementary material could be recast in terms of the signature arrival or transit times. Such a reanalysis could be useful for constraining arrival times in MHD models of ICME propagation such as ENLIL (Odstrcil, 2003), or for constraining models that use STEREO's *Heliospheric Imager* observations to predict ICME arrival times (*e.g.*, Möstl et al., 2014; Rollett et al., 2014);
- Individual, multipoint-observed events from the catalogues could be analysed in greater detail. Although statistical averages of ICME characteristics are valuable, ICMEs show enough variability for case studies to be worthwhile. The supplementary material could be used to study longitudinal variations in the signatures of individual events (*e.g.*, Farrugia et al., 2011), or, where observing spacecraft were radially aligned, the radial evolution in events (*e.g.*, Good et al., 2015). 23 flux ropes were observed by spacecraft pairs where the spacecraft separation was less than 12° in longitude and greater than 0.1 AU in heliocentric distance. A number of these events were well observed by both of the spacecraft in question. The magnetic cloud presented in Section 4.3.2 is a prime example of an event that would merit further study.

**Acknowledgements**

We wish to thank the MESSENGER, *Venus Express*, STEREO, ACE, and SOHO instrument teams for providing the data used in this work, as well as the PDS:PPI, ESA PSA, IGPP UCLA, ASC and CDAW data archives for their distribution of the data. This work has been supported with funding provided by the UK Science and Technology Facilities Council and the European Union's Seventh Framework Programme for research, technological development, and demonstration under grant agreement No. 606692 [HELCATS]. The referee's carefully considered and constructive comments on the work are much appreciated by the authors. We also wish to thank Volker Bothmer and Dániel Vech for useful discussions of this work.

**Figures**

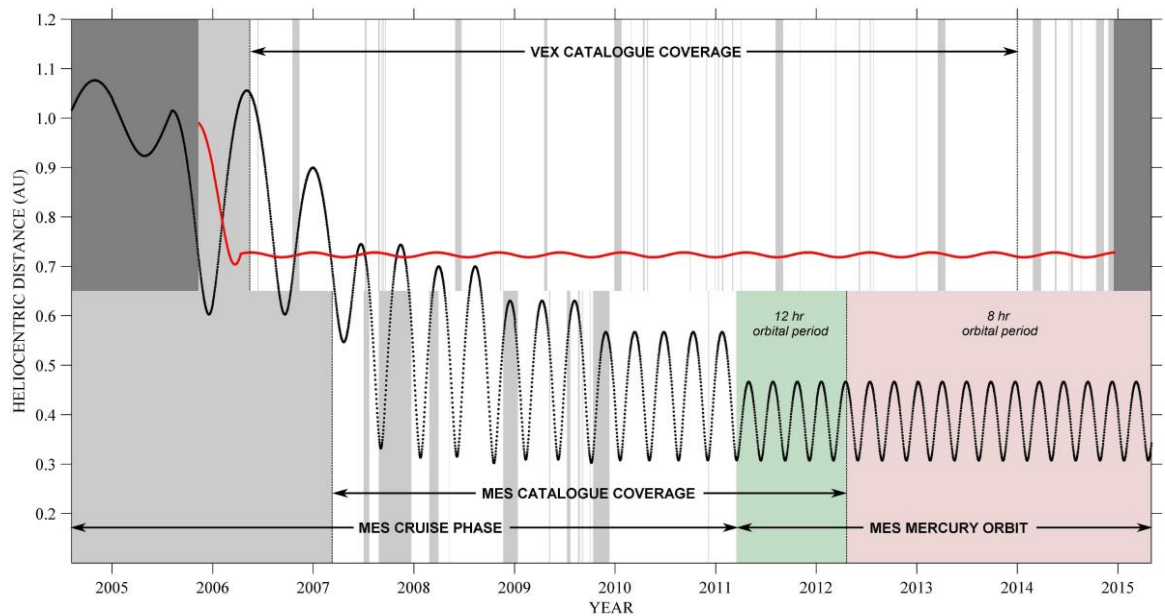

**Figure 1.** The heliocentric distances of MESSENGER (black-dotted line) and *Venus Express* (red line) with time. Vertical dotted lines bound the time period covered by the ICME catalogues. Regions shaded light grey in the top and bottom halves of the figure indicate times at which MAG data is unavailable at VEX and MES, respectively. The green-shaded region highlights the time when MES orbited Mercury with a period of 12 hours, and the red-shaded region with a period of eight hours.

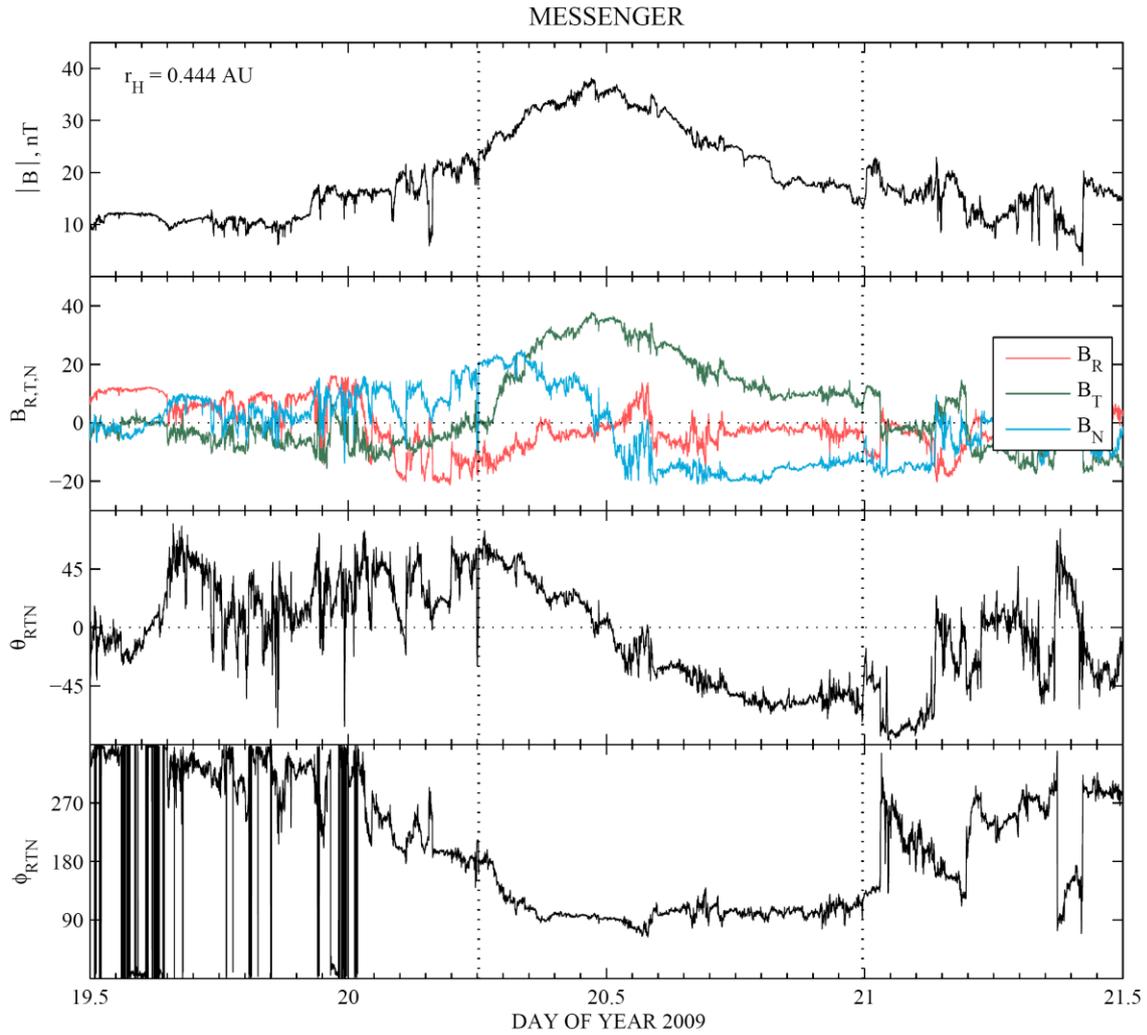

**Figure 2.** An ICME observed by MESSENGER during the spacecraft's cruise phase (event MES200901 in Table 1). The four panels show the B-field magnitude, RTN components and field angles $\theta_{RTN}$ and $\varphi_{RTN}$, respectively; the angles are as defined in the text. The flux rope is bounded by the dotted vertical lines, and displays an NWS configuration. The ICME is a quality 1 event.

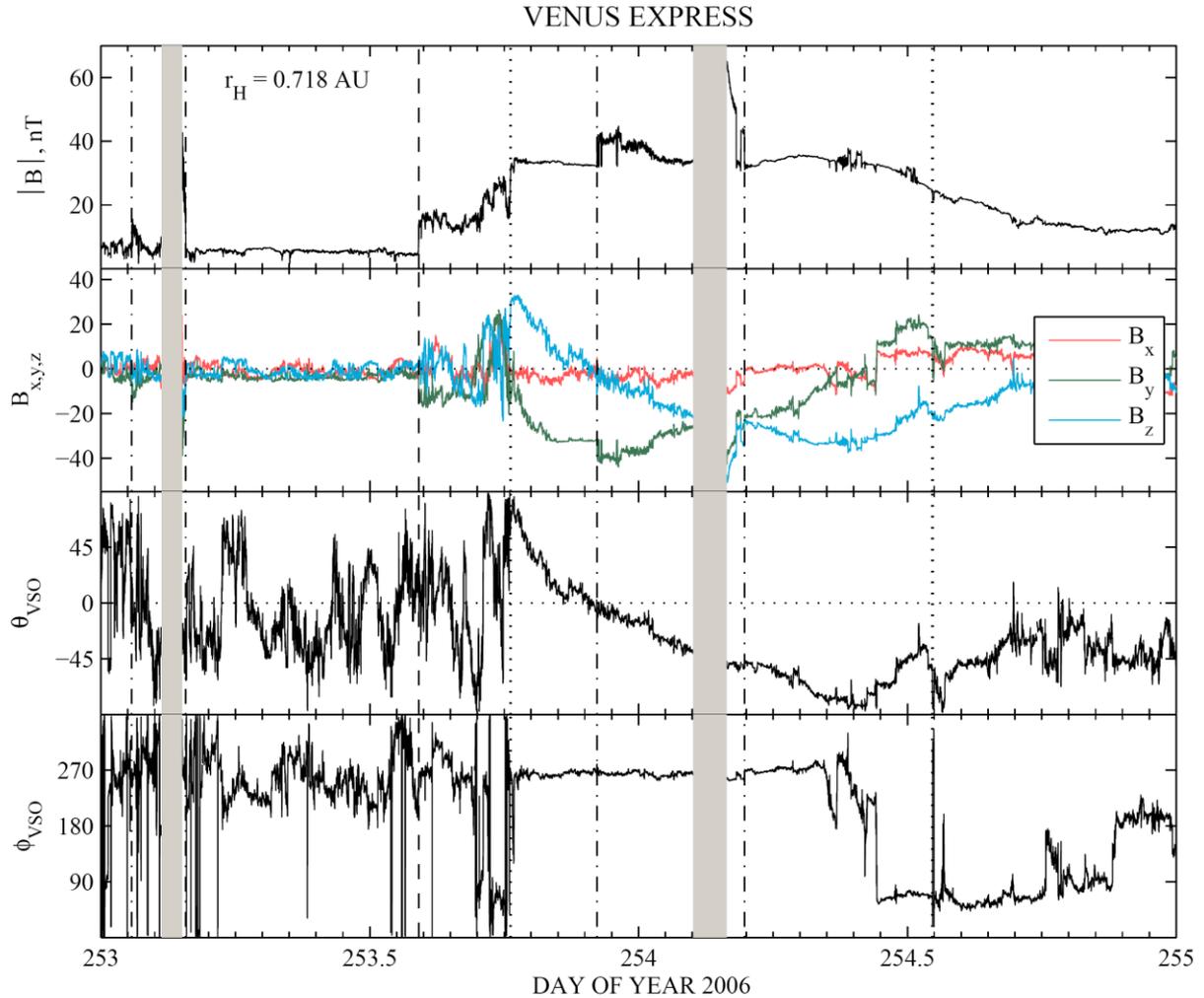

**Figure 3.** An ICME observed by *Venus Express* (event VEX200603). The panels show, respectively, the B-field magnitude, components and field angles in VSO co-ordinates; the angles are as defined in the text. Dotted lines show the flux rope boundaries, dash-dotted lines bow shock crossings, and the dashed line a field discontinuity driven by the rope; grey-shaded regions show periods when the spacecraft was close to or within Venus's induced magnetosphere. The ICME is a quality 1 event, and has an NWS-configured rope.

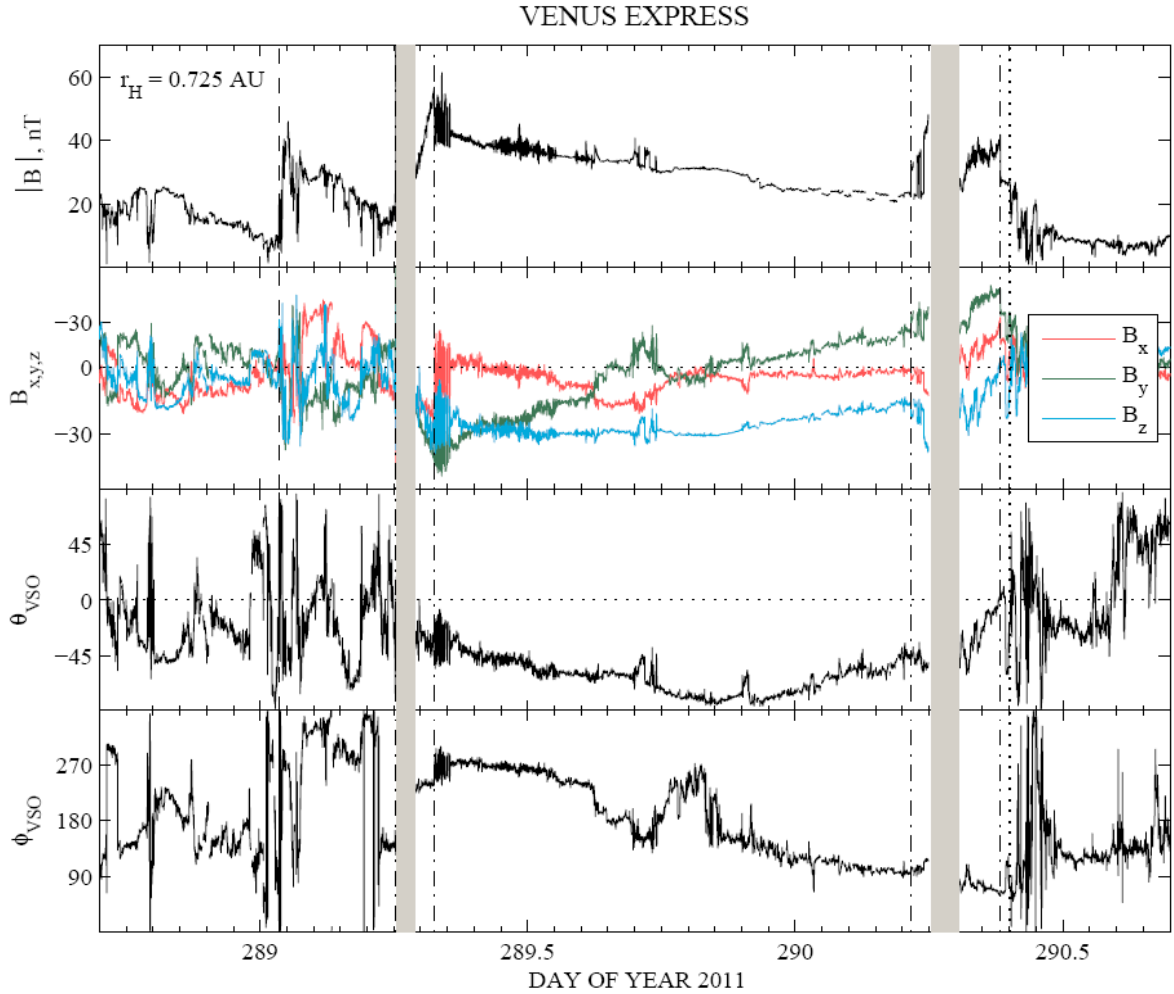

**Figure 4.** An ICME observed by *Venus Express* (event VEX201111). The panels and demarcation lines are as described in the caption to Figure 3. The ICME is a quality 2 event, with a WSE-configured rope.

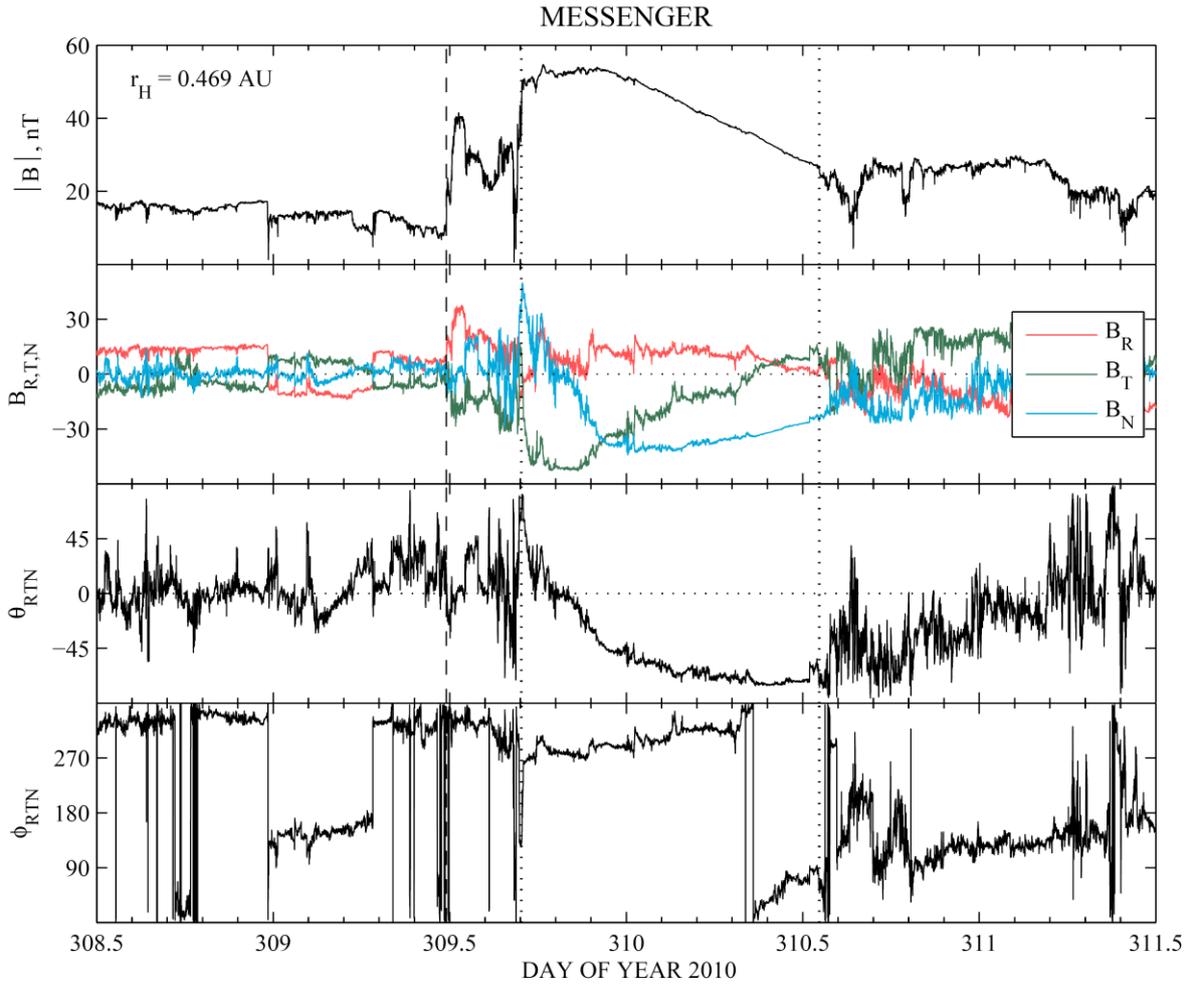

**Figure 5.** An ICME observed by MESSENGER (event MES201004) subsequently observed by STEREO-B (see Figure 6) while the spacecraft were close to radial alignment. The panels and demarcation lines are as described in previous figure captions.

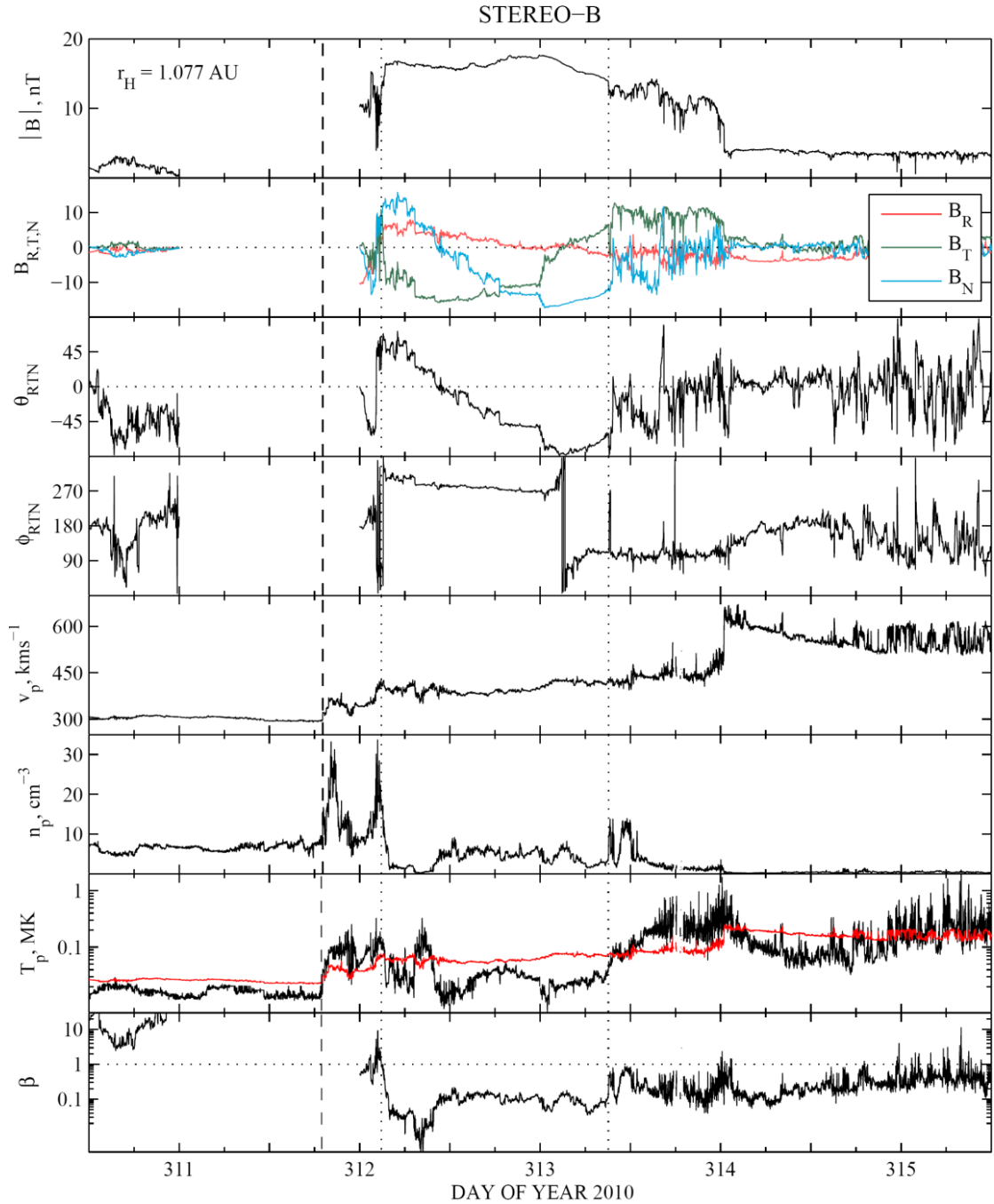

**Figure 6**: An ICME observed by STEREO-B, previously observed by MESSENGER (see Figure 5). The first four panels show magnetic field data in the same format as previous figures. The fifth to eighth panels show, respectively, the solar wind proton speed, proton density, proton temperature and plasma-beta. The red line overlaying the temperature data is the expected temperature (Lopez and Freeman, 1986), an empirically derived function of solar wind speed and heliocentric distance that predicts the temperature of normally-expanding solar wind. All of the magnetic field and plasma signatures of a magnetic cloud were present in this event. This cloud is clearly observed at both MES and STEREO-B.

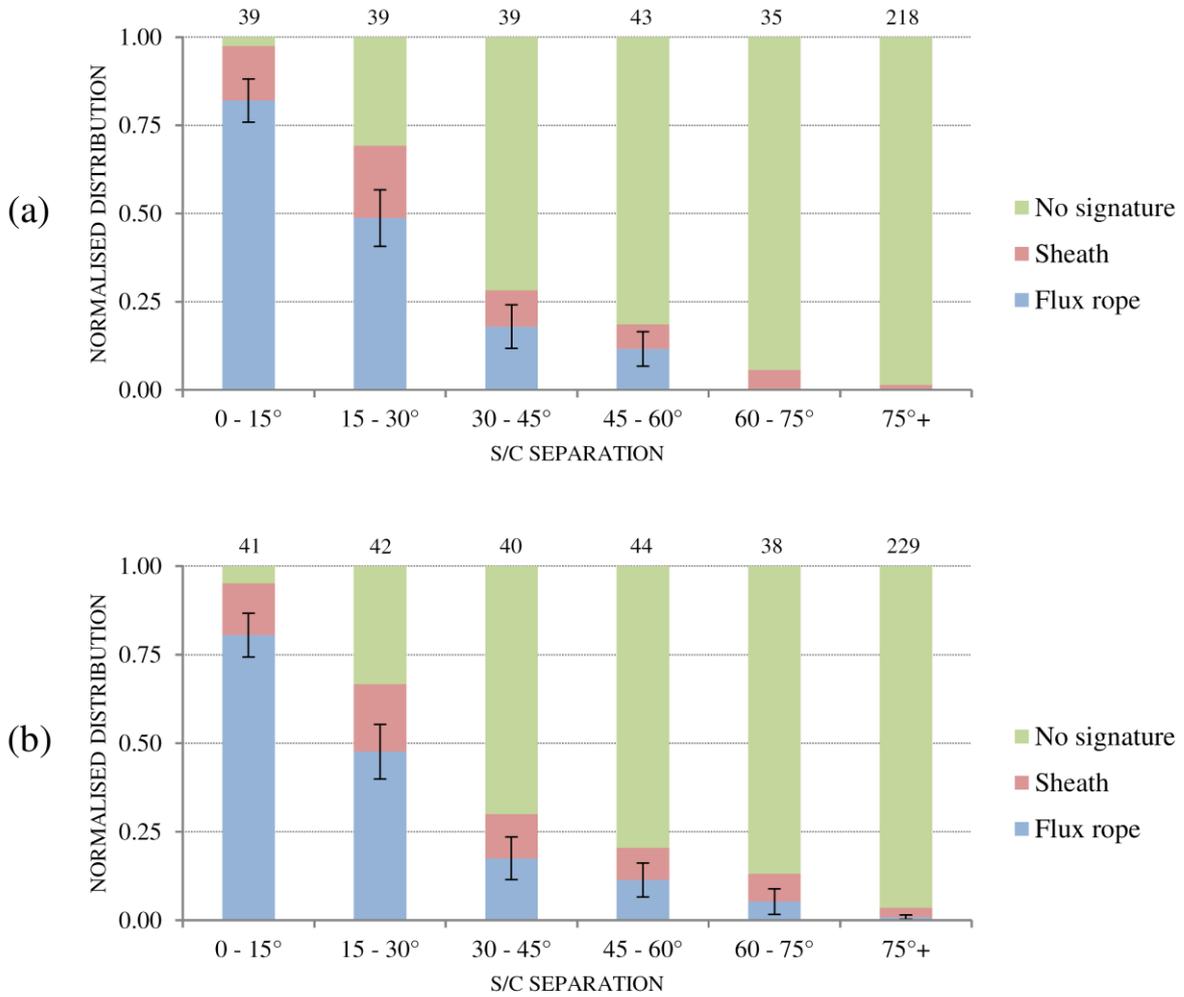

**Figure 7.** (a) Signatures of the ICMEs observed by MESSENGER and *Venus Express* at other spacecraft, as a function of spacecraft separation; (b) Same as (a), with the addition of less certain identifications.